\documentclass[11pt]{article}
\title{Flat coordinates and dilaton fields for three--dimensional conformal sigma models}

\def\ba{\begin{array}}
\def\ea{\end{array}}
\def\be{\begin{equation}}
\def\ee{\end{equation}}
\def\lbl{\label}

\def\eqn{equation}
\def\tfn{transformation}
\def\fn{function}
\def\-1{^{-1}}

\def\real{{\bf R}}
\def\cd{{\cal {D}}}
\def\cg{{\cal G}}
\def\tcg{\tilde{{\cal G}}}

\def\sm{$\sigma$--model}
\def\pl{Poisson--Lie}
\def\dd{Drinfel'd double}
\def\vbe{vanishing $\beta$ equations}

\def\email{e--mail: }
\author{Ladislav Hlavat\'y and Miroslav Turek \\
Faculty of Nuclear Sciences and Physical Engineering, \\
Czech Technical University\\
B\v rehov\'a 7, 115 19 Prague 1, Czech Republic\\
\email{hlavaty@fjfi.cvut.cz, turekm@km1.fjfi.cvut.cz}}
\begin{document}
\maketitle \abstract{Riemannian coordinates for flat metrics corresponding to three--dimensional conformal
Poisson--Lie T--dualizable sigma models are found by solving partial differential equations that follow from the
transformations of the connection components. They are then used for finding general forms of the dilaton fields
satisfying the vanishing beta equations of the sigma models.}

\section{Introduction} Recent interest in the conformal invariant \sm s follows from their relation to the
string theory. Principal \sm {} can be defined as a field theory on a Lie group $G$ on which a covariant second
order tensor field $F$ is given. The action of the \sm{} then is \be S_F[\phi]=\int d^2x \partial_-
\phi^{i}F_{ij}(\phi)
\partial_+ \phi^j \label{sigm1} \ee where the functions $ \phi^j:\ \real^2\rightarrow \real,\ j=1,2,\ldots,{\dim}\,G$ are obtained by the
composition $\phi^j=y^j\circ\phi \ $ of a map $
\phi:\real^2\rightarrow G$ and a coordinate map $y:U_g\rightarrow
\real^n,\ n={\dim}\,G$ of a neighborhood of an element
$\phi(x_+,x_-)=g\in G$.

The equations of motion have the form
\begin{equation}\label{eqmot}
 \partial_- \partial_+ \phi^{j}+\gamma_{rs}^j \partial_- \phi^{r}\partial_+
 \phi^{s}=0,
\end{equation}
where
\begin{equation}\label{chri}  \gamma_{rs}^j:=\frac{1}{2} G^{ji}(F_{is,r}+F_{ri,s}-F_{rs,i})
\end{equation}
and $G^{ji}$ is the inverse of \be G_{ij}=\frac{1}{2}(F_{ij}+F_{ji}).\label{gij} \ee

Quantization of the \sm s requires that they be made conformal invariant. This is achieved by addition of
another term depending on a scalar (dilaton) field $\Phi$ to the action (\ref{sigm1}). To guarantee the
conformal invariance of the \sm {} (at least at the one--loop level) the fields $F$ and $\Phi$ must satisfy the
so called \vbe
\begin{eqnarray}
\label{bt1} 0 & = & R_{ij}-\bigtriangledown_i\bigtriangledown_j\Phi- \frac{1}{4}H_{imn}H_j^{mn}, \\
 \label{bt2} 0 & = & \bigtriangledown^k\Phi H_{kij}+\bigtriangledown^k H_{kij},  \\
 \label{bt3} 0 & = & R-2\bigtriangledown_k\bigtriangledown^k\Phi- \bigtriangledown_k\Phi\bigtriangledown^k\Phi-
 \frac{1}{12}H_{kmn}H^{kmn},
 \end{eqnarray}
where covariant derivatives $\bigtriangledown_k$, Ricci tensor $R_{ij}$ and scalar curvature $R$ are calculated
from the (pseudo)metric (\ref{gij}) that is also used for lowering and raising indices. The components of
torsion are defined as \be H_{ijk}=\partial_i B_{jk}+\partial_j B_{ki}+\partial_k
B_{ij}, \lbl{torsion} \ee where %the $B$ is the so called torsion potential
\be B_{ij}=\frac{1}{2}(F_{ij}-F_{ji}).\lbl{torpot}\ee

We shall be interested in \sm s that satisfy the \vbe{} and moreover are \pl {} T--{} dualizable i.e. satisfy
\cite{kli:pltd}\be {\cal L}_{v_i}(F)_{\mu\nu}= F_{\mu\kappa}v^\kappa_j\tilde f_i^{jk}v^\lambda_kF_{\lambda\nu},
\ i=1,\ldots,\dim\, G, \label{klimseveq}\ee where $v_i$ form a basis of left--invariant fields on $G$ and
$\tilde f_i^{jk}$ are structure coefficients of a Lie group $\tilde G$ such that $\dim\,\tilde G=\dim\, G$. If
$F$ satisfies the \eqn {} (\ref{klimseveq}) then the equations of motion of the \sm{} can be rewritten (see
\cite{klse:dna}, \cite{kli:pltd}) as \eqn s for maps to the six--dimensional Drinfel'd double $D=(G|\tilde G)$
-- connected Lie group whose Lie algebra $\cal D$ admits a decomposition \be \cd =\cg + \tcg \label{decomp}\ee
into two subalgebras that are maximally isotropic with respect to a bilinear, symmetric, nondegenerate,
ad--invariant form on $\cd$.

There are two important types of coordinates on the manifolds where the \sm s live. The first one is given by
the Lie group structure and follows from the possibility to express the elements of the Lie group (at least in
the vicinity of the unit) as a product of elements of one--parametric subgroups. The Poisson--Lie T--dual \sm s
are usually expressed in terms of these group coordinates. The other type of coordinates are those in which the
metric on the manifold have a special simple form. They are called Riemannian coordinates (see e.g.
\cite{eisenhart:rg}). The Riemannian coordinates of the flat metrics will be called flat coordinates here. In
these coordinates the flat metric tensors become constant and the Christoffel symbols vanish so that the \eqn s
of motion (\ref{eqmot}) as well as the \vbe{} (\ref{bt1})--(\ref{bt3}) become very simple. That's why it is very
desirable to find the transformation between the group and Riemannian coordinates of the \sm s.
\section{Investigated models}\label{invmod} In the paper \cite{hlasno:3dsm2} the semiabelian Drinfel'd doubles
${(G|1)}$, for which $\cg$ in the decomposition (\ref{decomp}) are solvable Bianchi algebras ${\bf
2,3,4,5,6_0,7_0}$ and $\tcg$ is the three--dimensional Abelian Lie algebra ${\bf 1}$ (for the notation see
\cite{hlasno:3dsm2},\cite{Landau}), were  investigated and classification of conformal invariant \pl {}
T--dualizable \sm s with constant dilaton field was done. A bit surprisingly, all these models are not only
Ricci flat and torsionless but also flat in the sense that their Riemann--Christoffel tensor vanishes.

The metrics are expressed in the group coordinates $y_1,y_2,y_3$, for which elements of the group $G$ are
parametrized as (for typographic reasons we use subscripts for coordinates in the following)
\begin{equation}\label{groupelmt} g(y)=e^{y_{1}X_{1}}e^{y_{2}X_{2}}e^{y_{3}X_{3}},
\end{equation} where $X_{1},
X_{2}, X_{3}$ are generators of the corresponding Lie algebra, and we are going to find the \tfn s of the group
coordinates to the flat coordinates. The metrics corresponding to the investigated \dd s are
\begin{eqnarray}
\mbox{$(2|1):$}\nonumber\\
&G(y)_{ij}&= \left(\begin{array}{ccc}\label{F_{21}}
0 & u & v\\
u & q & g+uy_{2}\\
v & g+uy_{2} & r+2vy_{2}
\end{array}\right),\\
\mbox{$(3|1):$}\nonumber\\
&G(y)_{ij}&= \left(\begin{array}{ccc}\label{F_{31}}
p & u+ze^{-2y_{1}} & -u+ze^{-2y_{1}}\\
u+ze^{-2y_{1}}& q & -q\\
-u+ze^{-2y_{1}}& -q & q
\end{array}\right),\\ \nonumber\\
\mbox{$(4|1):$}\nonumber\\
&G(y)_{ij}&= \left(\begin{array}{ccc}\label{F_{41}}
p & (vy_{1}+u)e^{-y_{1}} & ve^{-y_{1}}\\
(vy_{1}+u)e^{-y_{1}} & qe^{-2y_{1}} & 0\\
ve^{-y_{1}} & 0 & 0
\end{array}\right),\\
%\end{eqnarray}
%\begin{eqnarray}
\mbox{$(5|1):$}\nonumber\\
&G(y)_{ij}&= \left(\begin{array}{ccc}\label{F_{51}}
p & ue^{-y_{1}} & ve^{-y_{1}} \\
ue^{-y_{1}} & \frac{g^2}{r}e^{-2y_{1}} & ge^{-2y_{1}}\\
ve^{-y_{1}} & ge^{-2y_{1}} & re^{-2y_{1}}
\end{array}\right),\\
\mbox{$(6_{0}|1):$}\nonumber\\
&G(y)_{ij}&= \left(\begin{array}{ccc}\label{F_{601}}
p & 0 & v+py_{2}\\
0 & -p & g-py_{1}\\
v+py_{2} & g-py_{1} & r+2gy_{1}+2vy_{2}+p(y_{2}^{2}-y_{1}^{2})
\end{array}\right),\nonumber\\ \\
\mbox{$(7_{0}|1):$}\nonumber\\
&G(y)_{ij}&= \left(\begin{array}{ccc}\label{F_{701}}
p & 0 & z+py_{2}\\
0 & p & g-py_{1}\\
z+py_{2} & g-py_{1} & r-2gy_{1}+2zy_{2}+p(y_{1}^{2}+y_{2}^{2})
\end{array}\right)\nonumber\\
\end{eqnarray}
where  $u,v,p,q,g,r,z$ are arbitrary real constants.

Beside these models, solutions of the \vbe{} with flat metrics  and nonconstant dilaton fields $\Phi$ were
found by the \pl{} T--duality %in \cite{hlasno:3dsm2}.
on the \dd {} $(1|6_{0})\cong(5ii|6_{0})\cong(1|6_0)$. The metrics and the dilaton fields expressed in the group
coordinates read
\begin{eqnarray}\label{F_{16}}\mbox{$(1|6_{0}):$}\nonumber\\
G(y)_{ij}&=& K(y_1,y_2)\-1\left(\begin{array}{ccc} -{{k}^{2} {q} {y_{1}}^{2}} &
{k}^{2} {q} {y_{1}} {y_{2}} & -{{k} ({1}+{k} {y_{1}})} \\
{{k}^{2} {q} {y_{1}} {y_{2}}} & {{q} (-{1}+{k}^{2} {y_{2}}^{2})} & {{k}^{2} {y_{2}}}\\
-{{k} ({1}+{k} {y_{1}})} & {{k}^{2} {y_{2}}} & 0
\end{array}\right),\\ \nonumber\\
%\Phi&=&\ln\left|(1+k(y_{1}-y_{2}))(1+k(y_{1}+y_{2}))\right|+C\label{phi11},\\
\Phi&=&\ln\left|(K(y_1,y_2)\right|+C\label{phi11},
\end{eqnarray}
where  $k,q$  are constants and $$K(y_1,y_2)={1}+{2} {k} {y_{1}}+{k}^{2} ({y_{1}}^{2}-{y_{2}}^{2}).$$
\begin{eqnarray}\label{F_{5ii60}}\mbox{$(5ii|6_{0}):$}\nonumber\\
G(y)_{11}&=&\frac{q\,(w^{2}-1)}{4\,W(y_1,y_2)}\ (1+e^{2y_{1}+2y_{2}}-2\,e^{2y_{1}+y_{2}})^2
,\nonumber\\
G(y)_{21}&=&\frac{q}{4\,W(y_1,y_2)}\
(1-2e^{2y_{1}+y_2}+e^{2y_{1}+y_{2}})\left(w^2 (1-2e^{y_{1}}+e^{2y_{1}+2y_{2}})-1-e^{2y_{1}+2y_{2}}\right),\nonumber\\
G(y)_{22}&=&\frac{q}{4\,W(y_1,y_2)}\ \left(w^2 (1-2e^{y_{1}}+e^{2y_{1}+2y_{2}})^2-(1+e^{2y_{1}+2y_{2}})^2\right),\nonumber\\
G(y)_{31}&=&\frac{w}{2\,W(y_1,y_2)}\ e^{y_{1}+y_{2}}
{\left((2e^{2y_{1}+y_{2}}-e^{2y_{1}+2y_{2}})(w-1)-w-1\right)}, \\
G(y)_{32}&=&\frac{w}{2\,W(y_1,y_2)} e^{y_{1}+y_{2}}\left({2w\,e^{y_{1}}-e^{2y_{1}+2y_{2}}(w-1)-w-1}\right),\nonumber\\
G(y)_{33}&=&0\nonumber
\end{eqnarray}
\begin{eqnarray}\Phi&=&\ln\left|(1+w)e^{-(y_{1}+y_{2})}+w(1-2e^{-y_{2}})
\right|+\ln\left|(w-1)e^{y_{1}+y_{2}}-w \right|+C,\label{phi22}
\end{eqnarray}
where  $w$ is a constant and $$W(y_1,y_2)=e^{y_{1}+y_{2}}{((w-1)\,e^{{y_{1}}+{y_{2}}}-w) ({1}+{w}-{2}w\,
e^{{y_{1}}} +w\,e^{{y_{1}}+{y_{2}}} )}.$$

The investigated models can also have nonzero antisymmetric part $B$ of the tensor $F$ but the corresponding
torsions $H_{ijk}$ given by (\ref{torsion}) are zero so that we assume that $F_{ij}=G_{ij}$ in the following. In
spite of the fact that all the metrics above are flat, the task to find coordinates for which the metrics become
constant is not trivial.
\section{Flat coordinates}
For finding the flat coordinates we shall use the formula for
transformation of the Levi--Civita connection
\begin{equation}
\Gamma^{i}_{jk}(y)=\frac{1}{2}G^{li}\Bigl( \frac{\partial
G_{kl}}{\partial y_{j}}+\frac{\partial G_{jl}}{\partial
y_{k}}-\frac{\partial G_{kj}}{\partial y_{l}}\Bigr).
\end{equation}
that reads as \begin{equation} \Gamma^{i}_{jk}(y)=\frac{\partial y_{i}}{\partial \xi^{l}}\frac{\partial
\xi_{m}}{\partial y_{j}}\frac{\partial \xi_{n}}{\partial y_{k}}\Gamma'^{l}_{mn}(\xi)+\frac{\partial
y_{i}}{\partial \xi_{l}}\frac{\partial^{2}\xi_{l}}{\partial y_{j}\partial y_{k}}.
\end{equation}
The components of $\Gamma'^{l}_{mn}(\xi)$ in the flat coordinates
vanish and we get the system of partial differential equations for
$\xi(y)$
\begin{equation}\label{rpns} \frac{\partial^{2} \xi_{i}}{\partial y_{j}\partial
y_{k}}=\Gamma^{l}_{jk}\frac{\partial \xi_{i}}{\partial y_{l}}.\\
\end{equation}
The system is linear and moreover separated with respect to the unknowns $\xi_i$'s. The possibility to solve it
explicitly depends on the form of $\Gamma^{l}_{jk}$. We were able to find general explicit solutions for the
metrics given above that together with the suitable initial conditions will produce the Riemannian coordinates.
The initial condition
\begin{equation}\label{initcon} \left[\frac{\partial
\xi_{k}}{\partial y_{i}}\right]_{\vec{y}=\vec{0}}=\delta^{i}_{k}
\end{equation}
produce the coordinates in which the metric acquires the constant form $\tilde G(\xi)=G(y={0})$ that can be
further diagonalized.

In the following we shall present solution of the equations
(\ref{rpns}) in detail for the metric (\ref{F_{601}})
 and write down the results for the other metrics.
\subsection{Flat coordinates for the \sm{} on  $(6_0|1)$}
The nonzero components of the affine connection for the metric
(\ref{F_{601}}) are
\begin{eqnarray}
\Gamma^{1}_{23}=1, &
\Gamma^{1}_{33}=\frac{-g+py_{1}}{p},\nonumber\\
\Gamma^{2}_{13}=1, &
\Gamma^{2}_{33}=\frac{v+py_{2}}{p},\nonumber\\
\end{eqnarray}
so that the equations (\ref{rpns}) read
\begin{eqnarray}\label{sdrr}
\frac{\partial^{2}\xi}{\partial y_{1}\partial y_{1}}&=&0,\label{r1}\\
\frac{\partial^{2}\xi}{\partial y_{1}\partial y_{2}}&=&0,\label{r2}\\
\frac{\partial^{2}\xi}{\partial y_{1}\partial y_{3}}&=&\frac{\partial\xi}{\partial y_{2}},\label{r3}\\
\frac{\partial^{2}\xi}{\partial y_{2}\partial y_{2}}&=&0,\label{r4}\\
\frac{\partial^{2}\xi}{\partial y_{2}\partial y_{3}}&=&\frac{\partial\xi}{\partial y_{1}},\label{r5}\\
\frac{\partial^{2}\xi}{\partial y_{3}\partial
y_{3}}&=&\left(\frac{-g+py_{1}}{p}\right)\frac{\partial\xi}{\partial
y_{1}}+\left(\frac{v+py_{2}}{p}\right)\frac{\partial\xi}{\partial
y_{2}}\label{r6}.
\end{eqnarray}
From (\ref{r1}) and (\ref{r2}) we get
\begin{eqnarray}\label{v1}
\xi=f(y_{3})\,y_{1}+h(y_{2},y_{3}).
\end{eqnarray}
and (\ref{r4}) and (\ref{r3}) imply
\begin{equation}\label{v3}
h(y_{2},y_{3})=f'(y_{3})\,y_{2}+b(y_{3}).
\end{equation}
The \eqn {} (\ref{r5}) gives
\be%gin{eqnarray}
%\frac{d^{2}f}{{dy_{3}}^{2}}-f=0\quad&\Rightarrow&\quad
f(y_{3})=ce^{y_{3}}+de^{-y_{3}}%\nonumber\\&\Rightarrow&\quad
%\xi=(ce^{y_{3}}+de^{-y_{3}})y_{1}+(ce^{y_{3}}-de^{-y_{3}})y_{2}+b(y_{3}).
\label{rr5}\ee
%\end{eqnarray}
and from (\ref{r6}) we get the \eqn{} for the \fn~$b$
\[
\frac{d^{2}b}{{dy_{3}}^{2}} =-\frac{g}{p}(ce^{y_{3}}+de^{-y_{3}})+\frac{v}{p}(ce^{y_{3}}-de^{-y_{3}})\] solved
by \begin{equation} b(y_{3})=-\frac{g}{p}(ce^{y_{3}}+de^{-y_{3}})+\frac{v}{p}(ce^{y_{3}}-de^{-y_{3}})+my_{3}+n
\end{equation}
so that the general solution of the system (\ref{r1})--(\ref{r6}) is
\begin{equation}
\xi(y_{1},y_{2},y_{3})=c(y_{1}+y_{2})e^{y_{3}}+d(y_{1}-y_{2})e^{-y_{3}}+\frac{c(v-g)}{p}e^{y_{3}}-\frac{d(v+g)}{p}e^{-y_{3}}+my_{3}+n,
\end{equation}
where $m, n, c, d$ are integration constants. As the transformation formulas (\ref{rpns}) are the same for all
the coordinate components $\xi_i$ we can write the flat coordinates in general as
\begin{eqnarray}
\xi_{1}(y_{1},y_{2},y_{3})&=&c_{1}(y_{1}+y_{2})e^{y_{3}}+d_{1}(y_{1}-y_{2})e^{-y_{3}}+\frac{c_{1}(v-g)}{p}e^{y_{3}}-\frac{d_{1}(v+g)}{p}e^{-y_{3}}\nonumber\\
&+&m_{1}y_{3}+n_{1},\nonumber\\
\xi_{2}(y_{1},y_{2},y_{3})&=&c_{2}(y_{1}+y_{2})e^{y_{3}}+d_{2}(y_{1}-y_{2})e^{-y_{3}}+\frac{c_{2}(v-g)}{p}e^{y_{3}}-\frac{d_{2}(v+g)}{p}e^{-y_{3}}\nonumber\\
&+&m_{2}y_{3}+n_{2},\nonumber\\
\xi_{3}(y_{1},y_{2},y_{3})&=&c_{3}(y_{1}+y_{2})e^{y_{3}}+d_{3}(y_{1}-y_{2})e^{-y_{3}}+\frac{c_{3}(v-g)}{p}e^{y_{3}}-\frac{d_{3}(v+g)}{p}e^{-y_{3}}\nonumber\\
&+&m_{3}y_{3}+n_{3}.\nonumber\\
\end{eqnarray}
and the integration constants will be determined by the required  form of the constant metric. When we choose \[
\left[\frac{\partial \xi_{k}}{\partial y_{i}}\right]_{\vec{y}=\vec{0}}=\delta^{k}_{i}\]then
\begin{eqnarray}
& &\xi_{1}(y_{1},y_{2},y_{3})=y_{1}\cosh(y_{3})+y_{2}\sinh(y_{3})+\frac{v}{p}\sinh(y_{3})-\frac{g}{p}\cosh(y_{3})-\frac{v}{p}y_{3}+n_{1},\nonumber\\
& &\xi_{2}(y_{1},y_{2},y_{3})=y_{1}\sinh(y_{3})+y_{2}\cosh(y_{3})+\frac{v}{p}\cosh(y_{3})-\frac{g}{p}\sinh(y_{3})+\frac{g}{p}y_{3}+n_{2},\nonumber\\
& &\xi_{3}(y_{1},y_{2},y_{3})=y_{3}+n_{3}.\nonumber\\
\end{eqnarray}
and\begin{equation}\label{g6con}
 \tilde G(\xi)= \left(\begin{array}{ccc}
p & 0 & v\\
0 & -p & g\\
v & g & r
\end{array}\right).
\end{equation} This constant form can be transformed by linear transformation
\begin{eqnarray}\label{yprime601}
y'_{1}&=&\left(\sqrt{\left|p\right|}\right)\xi_{1}+\varepsilon\left(\frac{v}{\sqrt{\left|p\right|}}\right)\xi_{3},\nonumber\\
y'_{2}&=&\left(\sqrt{\left|p\right|}\right)\xi_{2}-\varepsilon\left(\frac{g}{\sqrt{\left|p\right|}}\right)\xi_{3},\nonumber\\
y'_{3}&=&\left(\sqrt{\left|r+\left(\frac{\varepsilon
g^{2}}{|p|}-\frac{\varepsilon
v^{2}}{|p|}\right)\right|}\right)\xi_{3},
\end{eqnarray}
where
\begin{eqnarray*} &\varepsilon & = \left\{
\begin{array}{rl}
1 & \mbox{for}\ p>0\\
-1 & \mbox{for}\ p<0,
\end{array}
\right.\\ &\lambda & = \left\{
\begin{array}{rl}
1 & \mbox{for}\
\left(r+\frac{\varepsilon g^{2}}{|p|}-\frac{\varepsilon v^{2}}{|p|}\right)>0\\
-1 & \mbox{for}\ \left(r+\frac{\varepsilon g^{2}}{|p|}-\frac{\varepsilon v^{2}}{|p|}\right)<0
\end{array}
\right. \\
\end{eqnarray*}
(for $p=0$ or $r+\frac{g^{2}}{p}-\frac{v^{2}}{p}=0$ the metric is
not invertible) to the diagonal form
\begin{equation} {{G'}(y')}= \left(\begin{array}{ccc}
\varepsilon & 0 & 0\\
0 & -\varepsilon & 0\\
0 & 0 & \lambda
\end{array}\right).
\end{equation}

The solution of other cases is a bit more complicated, nevertheless, we were able to find the flat coordinates
in all investigated cases. Results are given below.
\subsection{Solution  for the \sm{} $(2|1)$}
The nonzero components of the affine connection for the metric
(\ref{F_{21}}) are
\begin{equation}
\begin{array}{rclrclrcl}
\Gamma^{1}_{22}&=&\frac{-u^{2}g-u^{3}y_{2}+uvq}{u^{2}r-2uvg+v^{2}q},&\Gamma^{1}_{23}&=&\frac{-vug-u^{2}vy_{2}+v^{2}q}{u^{2}r-2uvg+v^{2}q},&\Gamma^{1}_{33}&=&\frac{-vur-uv^{2}y_{2}+v^{2}g}{u^{2}r-2uvg+v^{2}q},\nonumber\\
\Gamma^{2}_{22}&=&\frac{-u^{2}v}{u^{2}r-2uvg+v^{2}q},&\Gamma^{2}_{23}&=&\frac{-uv^{2}}{u^{2}r-2uvg+v^{2}q},&\Gamma^{2}_{33}&=&\frac{-v^{3}}{u^{2}r-2uvg+v^{2}q},\nonumber\\
\Gamma^{3}_{22}&=&\frac{u^{3}}{u^{2}r-2uvg+v^{2}q},&\Gamma^{3}_{23}&=&\frac{u^{2}v}{u^{2}r-2uvg+v^{2}q},&\Gamma^{3}_{33}&=&\frac{uv^{2}}{u^{2}r-2uvg+v^{2}q}.\nonumber\\
\end{array}
\end{equation}
The general solution of the equations (\ref{rpns}) is
\begin{eqnarray}\xi(y_{1},y_{2},y_{3})&=&a-6d(u\rho-v\omega)^2y_1+b\,Y+c\,Y^2+d ( u\rho - v\omega)Y^3+\nonumber\\
& &(2c-6d\rho\omega)Z +3duv\,Z^2 -
     6dv\omega\,Y\,Z \end{eqnarray}
%     \be \,c(4) \left( 3\,u\,v\,Z^2 +\,Y^3\,\rho - v\,Y^3\,\omega  +
%     6\,v\,Y\,Z\,\omega  +  \right)+  Y^2\,c(5) - 2\,Z\,c(5) + Y\,c(6) + c(7)\ee
where $a, b, c, d$ are integration constants, \[ Y = u\,y_2+v\,y_3 ,\ \ Z=  \omega\,y_2+\rho\,y_3, \]
\[\omega=g\,u-q\,v,\ \ \rho=r\,u-g\,v.\] When we choose the initial
conditions (\ref{initcon}) then the flat coordinates in terms of the group coordinates are
\begin{eqnarray}
\xi_{1}(y_{1},y_{2},y_{3})&=&\frac{({6} {y_{1}} {u}^{2} {r}-{12} {y_{1}} {u} {v} {g}+{6} {y_{1}} {v}^{2} {q}-{3} {u}^{2} {y_{2}}^{2} {g}+{3} {u} {y_{2}}^{2} {v} {q}-{u}^{3} {y_{2}}^{3})}{6({u}^{2} {r}-{2} {u} {v} {g}+{v}^{2} {q})}\nonumber\\
& &+\frac{(-{3} {u}^{2} {v} {y_{3}} {y_{2}}^{2}-{3} {u} {y_{2}} {v}^{2} {y_{3}}^{2}-{6} {u} {y_{2}} {v} {y_{3}} {g}+{6} {y_{2}} {v}^{2} {y_{3}} {q}-{v}^{3} {y_{3}}^{3})}{6({u}^{2} {r}-{2} {u} {v} {g}+{v}^{2} {q})}\nonumber\\
& &+\frac{({v}^{2} {y_{3}}^{2} {g}-{v} {y_{3}}^{2} {u}
{r})}{2({u}^{2} {r}-{2} {u} {v} {g}+{v}^{2} {q})}+d_{1},\nonumber\\
\nonumber\\
\xi_{2}(y_{1},y_{2},y_{3})&=&\frac{(-{u}^{2} {v} {y_{2}}^{2}-{2} {u} {v}^{2} {y_{3}} {y_{2}}+{2} {r} {u}^{2} {y_{2}}-{4} {u} {v} {y_{2}} {g}+{2} {y_{2}} {v}^{2} {q}-{v}^{3} {y_{3}}^{2})}{2({u}^{2} {r}-{2} {u} {v} {g}+{v}^{2} {q})}+d_{2},\nonumber\\
\xi_{3}(y_{1},y_{2},y_{3})&=&\frac{({u}^{3} {y_{2}}^{2}+{2} {u}^{2} {v} {y_{3}} {y_{2}}+{u} {v}^{2} {y_{3}}^{2}+{2} {y_{3}} {u}^{2} {r}-{4} {v} {y_{3}} {u} {g}+{2} {v}^{2} {y_{3}} {q})}{2(u^{2}r-2uvg+v^{2}q)}+d_{3}.\nonumber\\
\end{eqnarray} and
\begin{equation}
{\widetilde G}(\xi)= \left(\begin{array}{ccc}
0 & u & v\\
u & q & g\\
v & g & r
\end{array}\right).
\end{equation}
By the linear transformation
\begin{eqnarray}
y'_{1}&=&\left(\frac{u}{\varepsilon\sqrt{\left|q\right|}}\right)\xi_{1}+\left(\sqrt{\left|q\right|}\right)\xi_{2}+\left(\frac{g}{\varepsilon\sqrt{\left|q\right|}}\right)\xi_{3},\nonumber\\
y'_{2}&=&\left(\frac{u}{\sqrt{\left|q\right|}}\right)\xi_{1}+\left(\left(\frac{ug}{q}-\frac{v}{\varepsilon}\right)\left(\frac{\sqrt{\left|q\right|}}{u}\right)\right)\xi_{3},\nonumber\\
y'_{3}&=&\sqrt{\left|\left(\frac{gu}{\left|q\right|}-\frac{v}{\varepsilon}\right)^2\frac{\left|q\right|}{u^{2}}-\frac{\varepsilon
g^{2}}{\left|q\right|}+\frac{r}{\varepsilon}\right|}\ \xi_{3},
\end{eqnarray}
where
\begin{eqnarray*} &\varepsilon &=\left\{
\begin{array}{rl}
1 & \mbox{for}\ q>0\\
-1 & \mbox{for}\ q<0,
\end{array}
\right.\\
&\lambda&=\left\{
\begin{array}{rl}
1 & \mbox{for}\ \left(\left(\frac{gu}{\left|q\right|}-\frac{v}{\varepsilon}\right)^2\frac{\left|q\right|}{u^{2}}-\frac{\varepsilon g^{2}}{\left|q\right|}+\frac{r}{\varepsilon}\right)>0\\
-1 & \mbox{for}\
\left(\left(\frac{gu}{\left|q\right|}-\frac{v}{\varepsilon}\right)^2\frac{\left|q\right|}{u^{2}}-\frac{\varepsilon
g^{2}}{\left|q\right|}+\frac{r}{\varepsilon}\right)<0
\end{array}
\right.
\end{eqnarray*}
we can transform the metric tensor (\ref{F_{21}}) to  the constant
diagonal form
\begin{equation}
{{G'}(y')}= \left(\begin{array}{ccc}
\varepsilon & 0 & 0\\
0 & -\varepsilon & 0\\
0 & 0 & \lambda\varepsilon
\end{array}\right).
\end{equation}
\subsection{Solution  for the \sm{} $(3|1)$}
The nonzero components of the affine connection for the metric
(\ref{F_{31}}) are
\begin{equation}
\begin{array}{rclrclrcl}
\Gamma^{1}_{11}&=&-2,&\Gamma^{2}_{11}&=&\frac{(pq-u^{2})e^{2y_{1}}+zu}{qz},&\Gamma^{3}_{11}&=&\frac{(pq-u^{2})e^{2y_{1}}-zu}{qz}.\\
\end{array}
\end{equation}
The general solution of the equations (\ref{rpns}) is
\begin{equation}
\xi(y_{1},y_{2},y_{3})=cy_{3}+ay_{2}+\frac{u(a-c)}{2q}y_{1}+\frac{(pq-u^{2})(a+c)}{8qz}e^{2y_{1}}+de^{-2y_{1}}+b,\\
\end{equation}
where $a, b, c, d$ are integration constants. When we choose the initial conditions (\ref{initcon}) then the
flat coordinates in terms of the group coordinates are
\begin{eqnarray}
\xi_{1}(y_{1},y_{2},y_{3}) &=& -\frac{1}{2}e^{-2y_{1}}+b_{1},\nonumber\\
\xi_{2}(y_{1},y_{2},y_{3}) &=& y_{2}+\frac{u}{2q}y_{1}+\frac{(pq-u^{2})}{8qz}e^{2y_{1}}+\frac{(pq-u^{2}+2uz)}{8qz}e^{-2y_{1}},\\
& & +b_{2},\\
\xi_{3}(y_{1},y_{2},y_{3}) &=& y_{3}-\frac{u}{2q}y_{1}+\frac{(pq-u^{2})}{8qz}e^{2y_{1}}+\frac{(pq-u^{2}-2uz)}{8qz}e^{-2y_{1}}\nonumber\\
& & +b_{3},\nonumber
\end{eqnarray}
and
\begin{equation}
{\widetilde G(\xi)}= \left(\begin{array}{ccc}
p & u+z & z-u\\
u+z & q & -q\\
z-u & -q & q
\end{array}\right).
\end{equation}
By the linear transformation
\begin{eqnarray}
y'_{1}&=&\left(\sqrt{\left|p\right|}\right)\xi_{1}+\left(\frac{u+z}{\varepsilon\sqrt{\left|p\right|}}\right)\xi_{2}+\left(\frac{z-u}{\varepsilon\sqrt{\left|p\right|}}\right)\xi_{3},\nonumber\\
y'_{2}&=&\left(\sqrt{\left|\frac{(u+z)^{2}}{\left|p\right|}-\frac{q}{\varepsilon}\right|}\right)\xi_{2}+\left(\frac{\frac{q}{\delta\varepsilon}+\frac{(z^{2}-u^{2})}{\delta\left|p\right|}}{\sqrt{\left|\frac{(u+z)^{2}}{\left|p\right|}-\frac{q}{\varepsilon}\right|}}\right)\xi_{3},\nonumber\\
y'_{3}&=&\left(\sqrt{\left|\frac{{\varepsilon\delta(\frac{(z^{2}-u^{2})}{\delta\left|p\right|}+\frac{q}{\delta\varepsilon})}^{2}}{(\frac{(u+z)^{2}}{\left|p\right|}-\frac{q}{\varepsilon})}-\frac{(z-u)^{2}}{\varepsilon\left|p\right|}+q\right|}\right)\xi_{3},
\end{eqnarray}
where
\begin{eqnarray*}
&\varepsilon &=\left\{
\begin{array}{rl}
1 & \mbox{for}\ p>0\\
-1 & \mbox{for}\ p<0,
\end{array}
\right.\\
&\delta&=\left\{
\begin{array}{rl}
1 & \mbox{for}\ \left(\frac{(u+z)^{2}}{\left|p\right|}-\frac{q}{\varepsilon}\right)>0\\
-1 & \mbox{for}\ \left(\frac{(u+z)^{2}}{\left|p\right|}-\frac{q}{\varepsilon}\right)<0,
\end{array}
\right.\\
&\lambda&=\left\{
\begin{array}{rl}
1 & \mbox{for}\ \left(\varepsilon\delta\frac{{(\frac{z^{2}-u^{2}}{\delta\left|p\right|}+\frac{q}{\delta\varepsilon})}^{2}}{(\frac{(u+z)^{2}}{\left|p\right|}-\frac{q}{\varepsilon})}-\frac{(z-u)^{2}}{\varepsilon\left|p\right|}+q\right)>0\\
-1 & \mbox{for}\
\left(\varepsilon\delta\frac{{(\frac{z^{2}-u^{2}}{\delta\left|p\right|}+\frac{q}{\delta\varepsilon})}^{2}}{(\frac{(u+z)^{2}}{\left|p\right|}-\frac{q}{\varepsilon})}-\frac{(z-u)^{2}}{\varepsilon\left|p\right|}+q\right)<0
\end{array}
\right.
\end{eqnarray*}
we can transform the metric tensor (\ref{F_{31}}) to  the constant
diagonal form
\begin{equation}
{{G'}(y')}= \left(\begin{array}{ccc}
\varepsilon & 0 & 0\\
0 & -\varepsilon\delta & 0\\
0 & 0 & \lambda
\end{array}\right).
\end{equation}
\subsection{Solution  for the \sm{} $(4|1)$}
The nonzero components of the affine connection for the metric
(\ref{F_{41}}) are
\begin{equation}
\begin{array}{rclrclrcl}
\Gamma^{1}_{11}&=&-1&\Gamma^{2}_{11}&=&\frac{v}{q}e^{y_{1}}&\Gamma^{2}_{12}&=&-1\\
\Gamma^{2}_{21}&=&-1&\Gamma^{3}_{11}&=&(\frac{p}{v}-\frac{u}{q}-\frac{v}{q}\,y_{1})e^{y_{1}}&\Gamma^{3}_{12}&=&\frac{u}{v}+y_{1}\\
\Gamma^{3}_{22}&=&\frac{q}{v}e^{-y_{1}}\\
\end{array}
\end{equation}
The general solution of the equations (\ref{rpns}) is
\begin{eqnarray}
\xi(y_{1},y_{2},y_{3}) &=&
cy_{3}+\frac{qc}{2v}y_{2}^{2}e^{-y1}+ay_{2}e^{-y_{1}}+cy_{1}y_{2}+\frac{cu}{v}y_{2}-cy_{2}+\frac{av}{q}y_{1}\nonumber\\
& & +\frac{pc}{2v}e^{y_{1}}+\frac{cv}{2q}e^{y_{1}}+de^{-y_{1}}+b,
\end{eqnarray}
where $a, b, c, d$ are integration constants. When we choose the initial conditions (\ref{initcon}) then the
flat coordinates in terms of the group coordinates are
\begin{eqnarray}
\xi_{1}(y_{1},y_{2},y_{3}) &=& -e^{-y_{1}}\nonumber\\
\xi_{2}(y_{1},y_{2},y_{3}) &=& y_{2}e^{-y_{1}}+\frac{v}{q}y_{1}+\frac{v}{q}e^{-y_{1}}\nonumber\\
\xi_{3}(y_{1},y_{2},y_{3}) &=& \frac{(pq-2uv)}{2qv}e^{-y_{1}}+\frac{v}{2q}e^{-y_{1}}+y_{3}+\frac{q}{2v}y_{2}^{2}e^{-y_{1}}+y_{1}y_{2}+\frac{u}{v}y_{2}\nonumber\\
& & -y_{2}+\frac{p}{2v}e^{y_{1}}-\frac{u}{v}y_{2}e^{-y_{1}}+y_{2}e^{-y_{1}}+\frac{(v-u)}{q}y_{1}-\frac{v}{2q}e^{y_{1}}.\nonumber\\
\end{eqnarray}
and
\begin{equation}
{\widetilde G(\xi)}= \left(\begin{array}{ccc}
p & u & v\\
u & q & 0\\
v & 0 & 0
\end{array}\right).
\end{equation}
By the linear transformation
\begin{eqnarray}
y'_{1}&=&\sqrt{|q|}\xi_{2}+\frac{\varepsilon u}{\sqrt{|q|}}\xi_{1},\nonumber\\
y'_{2}&=&\sqrt{\left|p-\frac{\varepsilon u^{2}}{|q|} \right|}\xi_{1}+\left(\frac{\delta v}{\sqrt{\left|p-\frac{\varepsilon u^{2}}{|q|} \right|}} \right)\xi_{3},\nonumber\\
y'_{3}&=&\left(\sqrt{\left|\frac{\delta
v^{2}}{\left|p-\frac{\varepsilon u^{2}}{|q|}\right|}\right|}
\right)\xi_{3},
\end{eqnarray}
where
\begin{eqnarray*}
&\varepsilon &=\left\{
\begin{array}{rl}
1 & \mbox{for}\ q>0\\
-1 & \mbox{for}\ q<0,
\end{array}
\right.\\
&\delta&=\left\{
\begin{array}{rl}
1 & \mbox{for}\ \left(p-\frac{\varepsilon u^{2}}{|q|} \right)>0\\
-1 & \mbox{for}\ \left(|p-\frac{\varepsilon u^{2}}{|q|} \right)<0,
\end{array}
\right.\\
&\lambda&=\left\{
\begin{array}{rl}
1 & \mbox{for}\ \left(\frac{\delta
v^{2}}{\left|p-\frac{\varepsilon u^{2}}{|q|}\right|}\right)>0\\
-1 & \mbox{for}\ \left(\frac{\delta v^{2}}{\left|p-\frac{\varepsilon u^{2}}{|q|}\right|}\right)<0
\end{array}
\right.
\end{eqnarray*}
we can transform the metric tensor (\ref{F_{41}}) to  the constant
diagonal form
\begin{equation}
{{G'}(y')}= \left(\begin{array}{ccc}
\varepsilon & 0 & 0\\
0 & \delta & 0\\
0 & 0 & -\lambda
\end{array}\right).
\end{equation}
\subsection{Solution  for the \sm{} $(5|1)$}
The nonzero components of the affine connection for the metric
(\ref{F_{51}}) are
\begin{equation}
\begin{array}{rclrclrcl}
\Gamma^{1}_{11}&=&-1&\Gamma^{2}_{11}&=&\frac{rp}{(ur-vg)}e^{y_{1}}&\Gamma^{2}_{12}&=&\frac{gv}{(ur-vg)}\nonumber\\
\Gamma^{2}_{13}&=&\frac{vr}{(ur-vg)}&\Gamma^{2}_{22}&=&\frac{g^{2}}{(ur-vg)}e^{-y_{1}}&\Gamma^{2}_{23}&=&\frac{gr}{(ur-vg)}e^{-y_{1}}\nonumber\\
\Gamma^{2}_{33}&=&\frac{r^{2}}{(ur-vg)}e^{-y_{1}}&\Gamma^{3}_{11}&=&-\frac{pg}{(ur-vg)}e^{y_{1}}&\Gamma^{3}_{21}&=&-\frac{ug}{(ur-vg)}\nonumber\\
\Gamma^{3}_{22}&=&-\frac{g^{3}}{r(ur-vg)}e^{-y_{1}}&\Gamma^{3}_{31}&=&-\frac{ur}{(ur-vg)}&\Gamma^{3}_{32}&=&-\frac{g^{2}}{(ur-vg)}e^{-y_{1}}\nonumber\\
\Gamma^{3}_{33}&=&-\frac{gr}{(ur-vg)}e^{-y_{1}}.
\end{array}
\end{equation}
The general solution of the equations (\ref{rpns}) is \be \frac{1}{2} a p\, e^{y_1} +a\, (u y_2+v y_3)+e^{-y_1}
\left(\frac{(g y_2+r
   y_3) (2 c r+a r y_3+a g y_2)}{2 r}-b\right)+ d,\ee
where $a, b, c, d$ are integration constants. When we choose the initial conditions (\ref{initcon}) then the
flat coordinates in terms of the group coordinates are
\begin{eqnarray}
\xi_{1}(y_{1},y_{2},y_{3})&=&-e^{-y_{1}}+d_{1}\nonumber\\
\xi_{2}(y_{1},y_{2},y_{3})&=&\frac{p\, r \cosh y_1+{r (u y_2+v y_3)}+{e^{-y_1} \left(\frac{1}{2} (g y_2+r
y_3)^2-v (g y_2+r
   y_3)\right)}}{r u-g v}+d_2\nonumber\\
\xi_{3}(y_{1},y_{2},y_{3})&=&\frac{-p\, g \cosh y_1+{g (u y_2+v y_3)}+{e^{-y_1} \left(-\frac{g}{2r} (g y_2+r
y_3)^2+u (g y_2+r
   y_3)\right)}}{r u-g v}+d_3\nonumber
\end{eqnarray}
and
\begin{equation}
{\widetilde G(\xi)}= \left(\begin{array}{ccc}
p & u & v\\
u & \frac{g^{2}}{r} & g\\
v & g & r
\end{array}\right).
\end{equation}
By the linear transformation
\begin{eqnarray}
y'_{1}&=&\left(\sqrt{\left|p\right|}\right)\xi_{1}+\left(\frac{u}{\varepsilon\sqrt{\left|p\right|}}\right)\xi_{2}+\left(\frac{v}{\varepsilon\sqrt{\left|p\right|}}\right)\xi_{3},\nonumber\\
y'_{2}&=&\left(\sqrt{\left|\frac{u^{2}}{\left|p\right|}-\frac{g^{2}}{\varepsilon r}\right|}\right)\xi_{2}+\left(\frac{\frac{vu}{\delta\left|p\right|}-\frac{g}{\delta\varepsilon}}{\sqrt{\left|\frac{u^{2}}{\left|p\right|}-\frac{g^{2}}{\varepsilon r}\right|}}\right)\xi_{3},\nonumber\\
y'_{3}&=&\left(\sqrt{\left|r-\frac{v^{2}}{\varepsilon\left|p\right|}+\frac{\varepsilon\delta\left(\frac{vu}{\left|p\right|}-g\varepsilon\right)^{2}}{\left|\frac{{u^2}}{\left|p\right|}-\frac{g^{2}}{\varepsilon
r}\right|}\right|}\right)\xi_{3},
\end{eqnarray}
where
\begin{eqnarray*}
&\varepsilon &=\left\{
\begin{array}{rl}
1 & \mbox{for}\ p>0\\
-1 & \mbox{for}\ p<0,
\end{array}
\right.\\
&\delta&=\left\{
\begin{array}{rl}
1 & \mbox{for}\ \left(\frac{u^{2}}{\left|p\right|}-\frac{g^{2}}{\varepsilon r}\right)>0\\
-1 & \mbox{for}\ \left(\frac{u^{2}}{\left|p\right|}-\frac{g^{2}}{\varepsilon r}\right)<0,
\end{array}
\right.\\
&\lambda&=\left\{
\begin{array}{rl}
1 & \mbox{for}\
\left(r-\frac{v^{2}}{\varepsilon\left|p\right|}+\frac{\varepsilon\delta\left(\frac{vu}{\left|p\right|}-g\varepsilon\right)^{2}}{\left|\frac{{u^2}}{\left|p\right|}-\frac{g^{2}}{\varepsilon
r}\right|}\right)>0\\
-1 & \mbox{for}\
\left(r-\frac{v^{2}}{\varepsilon\left|p\right|}+\frac{\varepsilon\delta\left(\frac{vu}{\left|p\right|}-g\varepsilon\right)^{2}}{\left|\frac{{u^2}}{\left|p\right|}-\frac{g^{2}}{\varepsilon
r}\right|}\right)<0
\end{array}
\right.
\end{eqnarray*}
we can transform the metric tensor (\ref{F_{51}}) to  the constant
diagonal form
\begin{equation}
{{G'}(y')}= \left(\begin{array}{ccc}
\varepsilon & 0 & 0\\
0 & -\delta\varepsilon & 0\\
0 & 0 & \lambda
\end{array}\right).
\end{equation}
\subsection{Solution  for the \sm{} $(7_0|1)$}
The nonzero components of the affine connection for this metric are
\begin{equation}
\begin{array}{rclrclrcl}
\Gamma^{1}_{23}&=&1&\Gamma^{1}_{32}&=&1&\Gamma^{1}_{33}&=&\frac{g}{p}-y_{1}\\
\Gamma^{2}_{13}&=&-1&\Gamma^{2}_{31}&=&-1&\Gamma^{2}_{33}&=&-\frac{z}{p}-y_{2}.
\end{array}
\end{equation}
The general solution of the equations (\ref{rpns}) is
\begin{equation}
\xi(y_{1},y_{2},y_{3})=
(\frac{g}{p}-y_{1})(iae^{iy_{3}}-ibe^{-iy_{3}})-(\frac{z}{p}+y_{2})(ae^{iy_{3}}+be^{-iy_{3}})+cy_{3}+d,
\end{equation}
where $a, b, c, d$ are integration constants. When we choose the initial conditions (\ref{initcon}) then the
flat coordinates in terms of the group coordinates are
\begin{eqnarray}
\xi_{1}(y_{1},y_{2},y_{3}) &=&(-\frac{g}{p}+y_{1})\cos(y_{3})+(\frac{z}{p}+y_{2})\sin(y_{3})-\frac{z}{p}y_{3}+d_{1}\nonumber\\
\xi_{2}(y_{1},y_{2},y_{3}) &=& (\frac{g}{p}-y_{1})\sin(y_{3})+(\frac{z}{p}+y_{2})\cos(y_{3})-\frac{g}{p}y_{3}+d_{2}\nonumber\\
\xi_{3}(y_{1},y_{2},y_{3}) &=& y_{3}+d_{3}.\nonumber\\
\end{eqnarray}
and
\begin{equation}\label{E_(7_{0}|1)}
{\widetilde G(\xi)}= \left(\begin{array}{ccc}
p & 0 & z\\
0 & p & g\\
z & g & r
\end{array}\right).
\end{equation}
By the linear transformation
\begin{eqnarray}
{\tilde y}_{1}&=&\left(\sqrt{\left|p\right|}\right)\xi_{1}+\frac{\varepsilon z}{\sqrt{\left|p\right|}}\xi_{3},\nonumber\\
{\tilde y}_{2}&=&\left(\sqrt{\left|p\right|}\right)\xi_{2}+\frac{\varepsilon g}{\sqrt{\left|p\right|}}\xi_{3},\nonumber\\
{\tilde y}_{3}&=&\left(\sqrt{\left|r-\frac{\varepsilon
z^{2}}{\left|p\right|}-\frac{\varepsilon
g^{2}}{\left|p\right|}\right|}\right)\xi_{3},
\end{eqnarray}
where
\begin{eqnarray*}
&\varepsilon &=\left\{
\begin{array}{rl}
1 & \mbox{for}\ p>0\\
-1 & \mbox{for}\ p<0,
\end{array}
\right.\\
&\lambda&=\left\{
\begin{array}{rl}
1 & \mbox{for}\ (r-\frac{\varepsilon z^{2}}{\left|p\right|}-\frac{\varepsilon
g^{2}}{\left|p\right|})>0\\
-1 & \mbox{for}\ (r-\frac{\varepsilon z^{2}}{\left|p\right|}-\frac{\varepsilon g^{2}}{\left|p\right|})<0
\end{array}
\right.
\end{eqnarray*}
we can transform the metric tensor (\ref{F_{701}}) to  the
constant diagonal form
\begin{equation}
{{G'}({\tilde y})}= \left(\begin{array}{ccc}
\varepsilon & 0 & 0\\
0 & \varepsilon & 0\\
0 & 0 & \lambda
\end{array}\right).
\end{equation}
\subsection{Solution  for the \sm{} $(1|6_0)$}
The nonzero components of the affine connection for the metric
(\ref{F_{16}}) are
\begin{equation}
\begin{array}{rclrclrcl}
\Gamma^{1}_{11}&=&-\frac{({1}+{k} {y_{1}}) {k}}{({1}+{2} {k}
{y_{1}}+{k}^{2} {y_{1}}^{2}-{k}^{2}
{y_{2}}^{2})}&\Gamma^{1}_{21}&=&\frac{{k}^{2} {y_{2}}}{({1}+{2}
{k} {y_{1}}+{k}^{2} {y_{1}}^{2}-{k}^{2}
{y_{2}}^{2})}\nonumber\\
\Gamma^{1}_{22}&=&-\frac{({1}+{k} {y_{1}}) {k}}{({1}+{2} {k}
{y_{1}}+{k}^{2} {y_{1}}^{2}-{k}^{2}
{y_{2}}^{2})}&\Gamma^{2}_{22}&=&\frac{{k}^{2} {y_{2}}}{({1}+{2}
{k} {y_{1}}+{k}^{2} {y_{1}}^{2}-{k}^{2} {y_{2}}^{2})}\nonumber\\
\Gamma^{2}_{11}&=&\frac{{k}^{2} {y_{2}}}{({1}+{2} {k}
{y_{1}}+{k}^{2} {y_{1}}^{2}-{k}^{2}
{y_{2}}^{2})}&\Gamma^{2}_{12}&=&-\frac{({1}+{k} {y_{1}})
{k}}{({1}+{2} {k} {y_{1}}+{k}^{2} {y_{1}}^{2}-{k}^{2}
{y_{2}}^{2})}\nonumber\\
\Gamma^{3}_{11}&=&\frac{({1}+{k} {y_{1}}) {k} {q}
{y_{1}}}{({1}+{2} {k} {y_{1}}+{k}^{2} {y_{1}}^{2}-{k}^{2}
{y_{2}}^{2})}&\Gamma^{3}_{21}&=&-\frac{{k}^{2} {q} {y_{1}} {y_{2}}}{({1}+{2} {k} {y_{1}}+{k}^{2} {y_{1}}^{2}-{k}^{2} {y_{2}}^{2})}\nonumber\\
\Gamma^{3}_{22}&=&\frac{{q} (-{1}+{k}^{2} {y_{2}}^{2}-{k}
{y_{1}})}{({1}+{2} {k} {y_{1}}+{k}^{2} {y_{1}}^{2}-{k}^{2}
{y_{2}}^{2})}
\end{array}
\end{equation}
The general solution of the equations (\ref{rpns}) is
\begin{eqnarray}
\xi(y_{1},y_{2},y_{3})&=&\frac{(qa+4kb)}{4k^{2}}\ln\left(1+k(y_{1}-y_{2})\right)+\frac{(qa+4kc)}{4k^{2}}\ln\left(1+k(y_{1}+y_{2})\right)\nonumber\\
& &
-\frac{qa}{2k}y_{1}+\frac{1}{4}\left(qa(y_{1}^{2}-y_{2}^{2})\right)+ay_{3}+d.
\end{eqnarray}
where $a, b, c, d$ are integration constants. When we choose the initial conditions (\ref{initcon}) then the
flat coordinates in terms of the group coordinates are
\begin{eqnarray}
\xi_{1}(y_{1},y_{2},y_{3})&=&\frac{1}{2k} {\ln}({1}+{k}
({y_{1}}-{y_{2}}))+\frac{1}{2k}{\ln}({1}+{k}
({y_{1}}+{y_{2}}))+d_{1}\nonumber\\
\xi_{2}(y_{1},y_{2},y_{3})&=&-\frac{1}{2k}{\ln}({1}+{k}
({y_{1}}-{y_{2}}))+\frac{1}{2k} {\ln}({1}+{k}
({y_{1}}+{y_{2}}))+d_{2}\nonumber\\
\xi_{3}(y_{1},y_{2},y_{3})&=&\frac{q}{4{k}^{2}}{\ln}({1}+{k}
({y_{1}}-{y_{2}}))+\frac{q}{4{k}^{2}}{\ln}({1}+{k}
({y_{1}}+{y_{2}}))-\frac{q}{2k}{y_{1}}\nonumber\\
& & +\frac{q}{4} ({y_{1}}^{2}-{y_{2}}^{2})
+{y_{3}}+d_{3}.\nonumber\\
\end{eqnarray}
and
\begin{equation}\label{E_(1|6)}
{\widetilde G(\xi)}= \left(\begin{array}{ccc}
0 & 0 & -k\\
0 & q & 0\\
-k & 0 & 0
\end{array}\right).
\end{equation}
By the linear transformation
\begin{eqnarray}\label{yy1}
y'_{1}&=&\left(\frac{1}{2}\sqrt{\left|2k\right|}\right)\xi_{1}+\left(\frac{1}{2}\sqrt{\left|2k\right|}\right)\xi_{3},\nonumber\\
y'_{2}&=&(\sqrt{\left|q\right|})\xi_{2},\nonumber\\
y'_{3}&=&\left(\frac{1}{2}\sqrt{\left|2k\right|}\right)\xi_{1}-\left(\frac{1}{2}\sqrt{\left|2k\right|}\right)\xi_{3},
\end{eqnarray}
where
\begin{eqnarray*} &\varepsilon &=\left\{
\begin{array}{rl}
1 & \mbox{for}\ k>0\\
-1 & \mbox{for}\ k<0,
\end{array}
\right.\\
&\delta&=\left\{
\begin{array}{rl}
1 & \mbox{for}\ q>0\\
-1 & \mbox{for}\ q<0
\end{array}
\right.
\end{eqnarray*}
we can transform the metric tensor (\ref{F_{16}}) to  the constant
diagonal form
\begin{equation}
{{G'}(y')}= \left(\begin{array}{ccc}
-\varepsilon & 0 & 0\\
0 & \delta & 0\\
0 & 0 & \varepsilon
\end{array}\right).
\end{equation}
\subsection{Solution  for the \sm{} $(5ii|6_0)$}
The general solution of the equations (\ref{rpns}) is
\begin{eqnarray}
\xi(y_{1},y_{2},y_{3})&=&-\frac{(2e^{y_{2}}qwa+(q(1+2w)a+2wb)y_{2}+e^{-(y_{1}+y_{2})}qa(2-w))}{4w^{2}}\nonumber\\
&-&\frac{(e^{y_{1}+y_{2}}qw^{2}a)}{4w^{2}}+\frac{(qa(w^{2}-1)+4wc)\ln(1+(-1+e^{-(y_{1}+y_{2})}))}{4w^{2}}\nonumber\\
&+&\frac{(qa(1+2w)+2wb)\ln(-2w+e^{y_{2}}w+e^{-y_{1}}(1+w))}{4w^{2}}+ay_{3}\nonumber\\
&+&\frac{(qwa+2wb)(y_{1}+y_{2})}{4w^{2}}+\frac{e^{y_{1}}qa(1+w)}{2w}+\frac{qa(y_{1}+y_{2})}{4w^{2}}+d\nonumber\\
\end{eqnarray}
where $a, b, c, d$ are integration constants. When we choose the initial conditions (\ref{initcon}) then the
flat coordinates in terms of the group coordinates are
\begin{eqnarray}
\xi_{1}(y_{1},y_{2},y_{3})=&-&\left(\frac{\ln(1-w+e^{-(y1+y2)}w)+\ln(-2we^{y_{1}}+e^{y_{1}+y_{2}}w+1+w)}{2w}\right)+d_{1}\nonumber\\
\xi_{2}(y_{1},y_{2},y_{3})=& &\left(\frac{\ln(-2we^{y_{1}}+e^{y_{1}+y_{2}}w+1+w)-\ln(1-w+e^{-(y_{1}+y_{2})}w)}{2w}\right)+d_{2}\nonumber\\
\xi_{3}(y_{1},y_{2},y_{3})=&-&\left(\frac{2qe^{-y_{2}}w^{2}+qwe^{-(y_{1}+y_{2})}-2qwe^{y_{1}}-qw^{2}e^{-(y_{1}+y_{2})}+qwe^{y_{1}+y_{2}}}{4w^{2}}\right)\nonumber\\
&+&\left(\frac{2e^{-y_{2}}qw^{2}-qw^{2}+e^{-(y_{1}+y_{2})}q\ln(1-w+e^{-(y_{1}+y_{2})}w)}{4w^{2}}\right)e^{y1+y2}\nonumber\\
&+&\frac{q}{4w^{2}}\ln(-2we^{-y_{2}}+w+e^{-(y_{1}+y_{2})}+e^{-(y_{1}+y_{2})}w)+\frac{q}{4w^{2}}(y_{2}+y_{1})\nonumber\\
&+&y_{3}+d_{3}
\nonumber\\
\end{eqnarray}
and
\begin{equation}\label{E_(5iii|6)}
{\widetilde G(\xi)}= \left(\begin{array}{ccc}
0 & 0 & w\\
0 & q & 0\\
w & 0 & 0
\end{array}\right).
\end{equation}
By the linear transformation
\begin{eqnarray}\label{yprime5ii60}
y'_{1}&=&\frac{1}{2}\left(\sqrt{2\left|w\right|}\right)\xi_{1}+\frac{1}{2}\left(\sqrt{2\left|w\right|}\right)\xi_{3},\nonumber\\
y'_{2}&=&(\sqrt{\left|q\right|})\xi_{2},\nonumber\\
y'_{3}&=&\frac{1}{2}\left(\sqrt{2\left|w\right|}\right)\xi_{1}-\frac{1}{2}\left(\sqrt{2\left|w\right|}\right)\xi_{3},
\end{eqnarray}
where
\begin{eqnarray*} &\varepsilon &=\left\{
\begin{array}{rl}
1 & \mbox{for}\ w>0\\
-1 & \mbox{for}\ w<0,
\end{array}
\right.\\
&\delta&=\left\{
\begin{array}{rl}
1 & \mbox{for}\ q>0\\
-1 & \mbox{for}\ q<0
\end{array}
\right.
\end{eqnarray*}
we can transform the metric tensor (\ref{F_{5ii60}}) to  the
constant diagonal form
\begin{equation}
{{G'}(y')}= \left(\begin{array}{ccc}
\varepsilon & 0 & 0\\
0 & \delta & 0\\
0 & 0 & -\varepsilon
\end{array}\right).
\end{equation}
\section{Dilaton fields}\label{s4}
As mentioned in the section \ref{invmod} the metrics (\ref{F_{21}}) -- (\ref{F_{701}}) were obtained in the
paper \cite{hlasno:3dsm2} from the requirement that the \vbe{} are satisfied for the constant dilaton field.
When we know the flat coordinates of these models we can easily find general forms of the dilaton fields that
together with these metrics satisfy the \vbe{} (\ref{bt1})--(\ref{bt3}) because in the flat coordinates these
\eqn s read
\begin{equation}\label{br22}
\frac{\partial^{2}\Phi'}{\partial y'_{i}\partial y'_{j}}= 0, \ \ \ G'^{ij}\frac{\partial\Phi'}{\partial
y'_{i}}\frac{\partial\Phi'}{\partial y'_{j}}=0.
\end{equation}
The general solution of these \eqn s is
\begin{equation}\label{phi}
\Phi'(y'_{1},y'_{2},y'_{3})=c_{1}y'_{1}+c_{2}y'_{2}+c_{3}y'_{3}+c_{4}\end{equation}where $G'^{ij}c_{i}c_{j}=0$.
%\begin{equation}G'_{11} c_{1}^{2}+G'_{22} c_{2}^{2}+G'_{33}c_{3}^{2}=0.\end{equation}

The general form of the dilaton field for the \sm {} $(6_0|1)$ with the metric (\ref{F_{601}}) that follow from
(\ref{phi}) and (\ref{yprime601}) is
\begin{eqnarray}
\Phi(y_{1},y_{2},y_{3})&=&c_{1}\left(\sqrt{\left|p\right|}\right)\left(y_{1}\cosh(y_{3})+y_{2}\sinh(y_{3})+\frac{v}{p}\sinh(y_{3})-\frac{g}{p}\cosh(y_{3})-\frac{v}{p}y_{3}\right)\nonumber\\
&+&c_{1}\left(\frac{\varepsilon
v}{\sqrt{\left|p\right|}}\right)y_{3}\nonumber\\
&+&c_{2}\left(\sqrt{\left|p\right|}\right)\left(y_{1}\sinh(y_{3})+y_{2}\cosh(y_{3})+\frac{v}{p}\cosh(y_{3})-\frac{g}{p}\sinh(y_{3})+\frac{g}{p}y_{3}\right)\nonumber\\
&-&c_{2}\left(\frac{\varepsilon
g}{\sqrt{\left|p\right|}}\right)y_{3}
+c_{3}\left(\sqrt{\left|\varepsilon
r+\left(\frac{g^{2}}{p}-\frac{v^{2}}{p}\right)\right|}\right)y_{3}+c_{4}\nonumber\\
\end{eqnarray}
where ${\rm sign}(\varepsilon)(c_{1}^{2}-c_{2}^{2})+{\rm sign}(\varepsilon\delta)c_{3}^{2}=0$.

By a similar way, i.e. as a linear combination of the flat coordinates, we can get the general dilaton fields
for the \sm s with the metrics (\ref{F_{21}})--(\ref{F_{51}}) and (\ref{F_{701}}). If the metric is positively
or negatively definite then the dilaton is constant.

We can  also get dilaton fields more general than (\ref{phi11}) and (\ref{phi22}) for the models $(1|6_0)$ and
$(5ii|6_0)$. The general form of the dilaton field  for the \sm {} $(1|6_0)$ is
\begin{eqnarray}\label{Phi11} \Phi(y)&=&\frac{\sqrt{|2k|}}{4k}\left(c_{1}+c_{3}+(c_{1}-c_{3})\frac{q}{2k}\right)
{\ln}\left|\left({1}+{k} ({y_{1}}-{y_{2}})\right)\left({1}+{k}
({y_{1}}+{y_{2}})\right)\right|\nonumber
\\&+&
(c_{1}-c_{3})\left[\frac{1}{2}\sqrt{\left|2k\right|}\left(-\frac{q}{2k}{y_{1}}
 +\frac{q}{4}({y_{1}}^{2}-{y_{2}}^{2})
+{y_{3}}\right)\right]\nonumber\\
&+&c_{2}\frac{\sqrt{\left|q\right|}}{2k}{\ln}\left|\frac{{1}+{k}
({y_{1}}+{y_{2}})}{{1}+{k} ({y_{1}}-{y_{2}})}\right|+c_{4}.
\end{eqnarray}
where ${\rm sign}(q)c_2^2+{\rm sign}(k)(c_3^2-c_1^2)=0$. For special choice of constants
$c_{1}=c_{3}={2k}/{\sqrt{|2k|}},\ c_{2}=0$, we get the dilaton field~(\ref{phi11}) obtained in
\cite{hlasno:3dsm2} by the \pl{} T--duality. The general form of the dilaton field for the \sm {} $(5ii|6_0)$
can be obtained from (\ref{yprime5ii60}) as well but it is rather extensive to display.

\section{Conclusions} We have obtained the explicit transformation between the group
coordinates of three--dimensional conformal \sm s living in the flat background and its Riemannian coordinates.
The forms of the metric in terms of the group coordinates were found in \cite{hlasno:3dsm2} from requirement of
conformal invariance of \sm s on the solvable Lie groups. The \tfn s were found by solving the \eqn s that
follow from the \tfn {} properties of the Levi--Civita connections. The results can be used for many purposes.
Let us mention two of them.

The \eqn s for the dilaton field of the flat \sm {} are easily solvable in the flat coordinates. In the Section
\ref{s4} we have shown that the coordinate transformations enable us to get the dilaton fields in terms of the
group coordinates. These, on the other hand, are convenient for obtaining the dilatons of the conformal \sm s
with nontrivial backgrounds by the \pl {} plurality \cite{unge}.

Besides that, the fact that all the investigated metrics are flat means that \eqn s of motion of the \sm s are
easily solvable in terms of the flat coordinates and the solutions can be transformed to the group coordinates.
The investigated models are \pl {} T--dual or plural to \sm s with nontrivial backgrounds and it offers a
possibility to find classical solutions in the nontrivial background. An example of such solution was given in
%( for $(2|1)\rightarrow (1|2) $
\cite{hla:slnbytduality} and other models are being solved now.

\end{document}